
\magnification=1100
\raggedbottom

\rightline{ Report Number RU93-3-B}
\vskip1cm

\centerline{\bf AN ALTERNATIVE PERTURBATIVE EXPANSION IN QUANTUM MECHANICS.}
\smallskip

\centerline{{\bf SCALING AND CUT-OFF RESUMMATION}
\footnote*{{\sevenrm Work supported in part by the U.S. Department of
Energy under Grant No. DE-91ER402651.B}}}
\vskip2.3cm

\centerline{J.M. PRATS}
\smallskip
\centerline{\it Department of Physics}
\centerline{\it The Rockefeller University}
\centerline{\it 1230 York Avenue, New York, N.Y. 10021}
\vskip3.5cm

\centerline{\bf Abstract}
\bigskip
An alternative perturbative expansion in quantum mechanics which allows
a full expression of the scaling arbitrariness is introduced. This
expansion is examined in the case of the anharmonic oscillator and
is conveniently resummed using a method which consists in introducing
an energy cut-off that is carefully removed as the order of the expansion
is increased. We illustrate this technique numerically by computing the
asymptotic behavior of the ground state energy of the anharmonic oscillator
for large couplings, and show how the exploitation of the scaling
arbitrariness substantially improves the convergence of this perturbative
expansion.
\vfill\eject

\noindent{\bf 1. Introduction}
\bigskip

Perturbation theory, as everybody knows, consists of calculating the
quantities associated to a Hamiltonian $H(g)$ as Taylor expansions in
the coupling $g$ around $g=0$, which corresponds to an exactly solvable
Hamiltonian. In systems like the anharmonic oscillator, however, these
series fail to converge. This problem is generally addressed by resumming
the series with the Pad\'e or Borel methods. In cases like the double
well potential or physically interesting field theories like QCD, however,
these methods are inapplicable due to the existance of instantons.
\smallskip

In this paper, we alternatively look at perturbation theory in a different,
very physical way. The central idea is the realization that all the
energy scales of the unperturbed Hamiltonian participate in the calculation
of an observable (an eigenstate, for instance) of the perturbed system at
a certain energy. This view naturally leads us to investigate two aspects of
perturbation theory. On one hand we investigate how every energy scale
contributes to the perturbed eigenstate and how the calculation of these
contributions can be optimized using energy-dependent scaling
transformations. This leads us to introduce a perturbative expansion that
exploits the scaling arbitrariness to a maximum.
On the other hand, we identify the divergences in the anharmonic oscillator
as due to the fact that the contributions calculated in perturbation
theory from arbitrarily high energies are unbounded. From this realization
we propose a resummation technique based on the introduction of an energy
cut-off.
\medskip

The optimization of perturbation theory using scaling properties is
accomplished in Quantum Field Theory (QFT) by the Renormalization
Group (RG)${}^{1,2}$. In the process of renormalization a new mass scale $\mu$
needs to be introduced and we get vertex functions expressed in terms
of more parameters than the physically independent ones. This parameter
redundancy is exploited by the RG: we can suitably change the scale
$\mu$ and the rest of parameters in such a way that we always remain in
the same physical theory. Although these changes do not affect the theory,
they do modify its perturbation expansions order by order. Thus, depending
on the energy scale at which we are working, we can select the appropriate
$\mu$ scale which optimizes this perturbative expansion. As is well known,
this technique has proved to be extremely fruitful in
QFT, its most important achievement being, perhaps, the discovery of
asymptotic freedom in non Abelian Gauge Theories${}^{3,4}$.
\smallskip

However, it is not clear to us that the conventional use of the RG fully
exploits the scaling properties of the theory. In the computation of vertex
functions at a certain energy, we will obviously use a $\mu$ scale of that
same energy. In that computation, however, all scales are involved since we
perform loop integrals over the full range of virtual momenta; the effect of
the perturbation on these other scales should, nevertheless, be optimally
computed using a different $\mu$.
\smallskip

In a recent paper${}^5$, we showed how the RG in a 4-dimensional $\phi^4$
theory could be extended to provide improved perturbation expansions
free of infrared divergences in the massless case. There, we computed
higher order vertex functions ($n>4$) using skeleton expansions in which
full propagators and 4-point functions are inserted respectively at
each line and vertex of the skeleton diagrams. In this way, depending
on the momenta running through a line or reaching a vertex of a diagram, the
propagators and couplings conveniently adjust themselves to those
particular scales. This would take care of the objection of the previous
paragraph for higher order vertex functions. Unfortunately, the equivalent
treatment for the primitive divergences would imply solving the
Schwinger-Dyson integral equations${}^6$, which is a highly non-trivial
matter.
\smallskip

In this paper we provide, for the simple case of one-dimensional
Field Theory (i.e., Quantum Mechanics), a perturbative expansion
that maximally exploits the arbitrariness of scaling in the sense
stated above. That is, we describe a perturbative expansion in which, in the
computation at a certain energy, the effect of the interaction on all the
other energies involved is suitably computed using the optimal scale for
every energy.
\medskip

Consider the Hamiltonian $H(g)=P^2/2m + V(g,x)$ where $g=0$ corresponds
to the exactly solvable system and $g=g_f$ to the perturbed one. We
shall consider the eigenvectors and eigenvalues of $H(g)$ as functions
of $g$. However, instead of doing the usual Taylor expansion in $g$,
we will use another expansion which is much more convenient for
visualizing how the various energy scales contribute to the calculation
and for exploiting the scaling arbitrariness. We will divide the interval
$g\in [0,g_f]$ into $N$ equal segments by introducing $N-1$ intermediate
points and will transport all eigenvectors and eigenvalues from
one point to the next using first order perturbation theory.
$N$ is here the order of our expansion; when $N$ is increased, the
intermediate segments become smaller and smaller and first order of
perturbation more and more accurate.
\smallskip

In this framework, scaling can be
easily implemented to improve perturbation theory at every step for each
energy level. If $g_1$ and $g_2$ are the initial and final points of a
certain elementary step, for every given energy level we appropriately
rescale $H(g_2)$ ($U^{-1}(\lambda)H(g_2)U(\lambda)=\lambda^2 \tilde H(g_2)$,
where $U(\lambda)$ is the scaling operator given in coordinate representation
by $U(\lambda)\Psi(x)=\sqrt{\lambda}\Psi(\lambda x)$) in such a way that
$H(g_1)$ and $\tilde H(g_2)$ are as similar as possible {\it at that
particular energy}. This requirement can be expressed by demanding that the
classical turning points of the two potentials coincide at that energy, since
this ensures that the corresponding wave functions will be essentially
confined to the same space interval, minimizing their difference.
We then compute the eigenvector $\tilde\Psi$ of $\tilde H(g_2)$
using first order perturbation theory and finally determine the
eigenvector $\Psi$ of $H(g_2)$ applying the scaling operator to
$\tilde\Psi$ ($\Psi=U(\lambda)\tilde\Psi$).
\medskip

We are essentially dividing the interaction introduced at each step
into two parts: an interaction that does not change the natural scale
of the eigenvectors and which is computed in first order of perturbation
theory, and a pure change of scale which is computed exactly by applying
the appropriate scaling transformation. In this way, we remove from
perturbation theory the burden of having to express the changes of scale
introduced by the interaction, and it can thus be more accurate. It is
important to recognize that these changes of scale introduced by the
interaction strongly depend on the energy and, therefore, the scaling
transformation $U(\lambda)$ will vary from eigenvector to eigenvector.
\medskip

Having introduced our alternative expansion and its optimization
by scaling in section 2, we turn in section 3 to the discussion of the
divergences of perturbation theory, particularizing for the anharmonic
oscillator.
\smallskip

As is well known, the divergence of the Taylor expansions for the
energy levels of the anharmonic oscillator is attributed to the fact
that for negative couplings the potential is unbounded from below
and the system becomes unstable. This instability manifests itself
in the form of singularities in the second and third sheets of the
Riemann surface associated to the analytic continuation of the
eigenvalues for complex couplings${}^{7,8}$. These singularities have $g=0$
as accumulation point and hence, the perturbation expansions for the
energy levels have a null radius of convergence. The knowledge of the
analytic properties of the eigenvalues${}^{7,8}$ and of the asymptotic
behavior of the Taylor coefficients${}^{9}$ enables us to use the powerful
machinery of the Borel Transformation${}^{10,11}$ to resum the series.
\medskip

In the framework of the perturbation expansion presented in this paper,
we look at the divergence in a different way. For large $N$, one would
expect that the errors made by first order of perturbation theory in an
elementary step are of order $(\Delta g)^2=(g_f/N)^2$; thus, the total
error for the $N$ steps would go like $1/N$ making the series
converge to the right result. However, this is not true for
the anharmonic oscillator. In this case, the perturbation changes the
asymptotic behavior of the potential for large $x$ and, consequently, the
perturbation becomes arbitrarily large for high energy states (which are
sensitive to the potential at large $x$). Thus, the errors made by first
order of perturbation theory in an elementary step cannot be uniformly
bounded with respect to energy by increasing $N$. The divergence in the low
energy states then arises because these big errors at high energy are
propagated down to the lower energy states in the subsequent elementary steps.
\smallskip

The use of the optimizing rescalings explained above ameliorates
this problem reducing substantially the errors at high energy. Nevertheless,
although the difference between the perturbed and unperturbed potentials
effective at each energy is greatly reduced by making their classical
turning points coincide, this difference is still arbitrarily large at
high energies and the divergence persists.
\smallskip

We show how this problem can be overcome introducing an energy cut-off by,
for example, limiting the number of states of the system. In this
way, we can reduce the errors of first order of perturbation theory
for {\it all} states as much as we want by increasing the number
of elementary steps. The theory can then be resummed by carefully
removing this cut-off as we increase the order of perturbation.
\smallskip

Thus, we regard our technique as a double expansion: in the number
of elementary steps (the order) and in the number of eigenstates
that we take around the energy at which we are computing (the cut-off).
The way in which we calculate the double limit of this expansion is
crucial: it can lead to the correct result or to a divergence. We
provide a criterion to find the suitable order of perturbation
associated to every value of the cut-off and show how, as expected,
this order is substantially lower for the expansion that fully
exploits the scaling arbitrariness.
\bigskip\bigskip\bigskip\bigskip

\noindent{\bf 2. An alternative perturbative expansion and scaling}
\bigskip

In this section we will introduce the mentioned alternative expansion
and will show how to fully exploit the scaling arbitrariness to
optimize it. We have in mind its application to the anharmonic oscillator,
but the formalism can be presented in a general way: we will consider a
Hamiltonian of the form $H(g)=P^2/2m + V(g,X)$ where the potential
satisfies $V(g,-x)=V(g,x),\quad V(g,x)<V(g,x')\ \ {\rm if} \ \
|x| < |x'| \quad {\rm and} \quad \lim_{|x|\to\infty}V(g,x)=\infty$.
\medskip

If $\Psi_n(g)$, $E_n(g)$ are respectively the eigenvectors and
eigenvalues of $H(g)$, we will first determine the expressions of
their derivatives with respect to $g$ in terms of them. The defining
equations are:

$$\eqalignno{H(g)\Psi_n(g)&=E_n(g)\Psi_n(g)&(1)\cr
(\Psi_n(g),\Psi_n(g))&=1&(2)\cr}$$

The eigenvectors are completely determined up to a global phase factor.
This phase can be conveniently chosen to satisfy the equation
$(\Psi_n(g),{\dot \Psi}_n(g))=0\quad(3)$. Differentiating (1) with
respect to $g$ yields:

$${\dot H}(g)\Psi_n(g)+H(g){\dot\Psi}_n(g)={\dot E}_n(g)\Psi_n(g)+
E_n(g){\dot\Psi}_n(g)\eqno(4)$$
\vfill\eject

And using (3) and the projection of (4) on $\Psi_k(g)$ $\forall k$,
we get te desired expressions:

$$\eqalignno{{\dot E}_n(g)&=(\Psi_n(g),{\dot H}(g)\Psi_n(g))&(5)\cr
{\dot\Psi}_n(g)&=\sum_{k\not= n}{(\Psi_k(g),{\dot H}(g)\Psi_n(g))
\over E_n(g)-E_k(g)}\Psi_k(g)&(6)\cr}$$

By repeatedly differentiating these equations with respect to $g$
and setting $g=0$, we obtain the ordinary perturbation theory. As we
pointed out in the introduction, we will use here a different strategy
which is much more convenient for later introducing the scaling
arbitrariness. We shall  alternatively expand in the number of intermediate
points in which we divide the interval $g\in[0,g_f]$ (the `order of
perturbation') and eigenvectors and eigenvalues will be propagated from
one point to the next using a first order approximation. In view of
(5)-(6), this first order approximation should be:

$$\eqalignno{\Psi_n(g+\Delta g)&\simeq\Psi_n(g)+\sum_{k\not= n}
{(\Psi_k(g),(H(g+\Delta g)-H(g))\Psi_n(g))\over E_n(g)-E_k(g)}
\Psi_k(g)&(7)\cr
E_n(g+\Delta g)&\simeq E_n(g)+(\Psi_n(g),(H(g+\Delta g)-H(g))
\Psi_n(g))&(8)\cr}$$

\noindent but it is, nevertheless, more convenient to replace
(7)-(8) by

$$\eqalignno{\Psi_n(g+\Delta g)&\simeq(\Psi_n(g)+\delta\Psi_n(g))/
\|\Psi_n(g)+\delta\Psi_n(g)\|&(7')\cr
{\rm where}\quad \delta\Psi_n(g)&=\sum_{k\not= n}{(\Psi_k(g),(H(g+\Delta g)
-H(g))\Psi_n(g)) \over E_n(g)-E_k(g)}\Psi_k(g)\cr
E_n(g+\Delta g)&\simeq(\Psi_n(g+\Delta g),H(g+\Delta g)\Psi_n(g+\Delta g))
&(8')\cr}$$

\noindent because, although these formulas coincide with (7)-(8) at first
order, they have the advantage of yielding exactly normalized states
and the exact energies corresponding to the (approximate) eigenstates for
finite values of $\Delta g$.
\medskip

No scaling properties have yet been introduced; so far we have just
changed the way in which we do our expansions. The different philosophies
underlying each of these two expansions, can probably be best illustrated
in the more familiar context of expanding a function around a point.
\smallskip

Suppose that we want to solve the equation

$${\partial g\over\partial x}(x,y)=O_yg(x,y)\eqno (9)$$

\noindent with the boundary condition $g(0,y)=f(y)$,
($O_y$ is an operator that acts on the $y$ variable only). Here are
the two strategies:
\smallskip

\noindent i) Differentiating (9) with respect to $x$ and setting $x=0$,
we get:

$${\partial^n g \over \partial x^n}(0,y)=O^n_y g(0,y)=O^n_y f(y)$$
\vfill\eject

\noindent and we can build the Taylor expansion

$$g(x,y)=\lim_{N\to\infty}\sum_{n=0}^N {x^n \over n!}O^n_y f(y)\eqno(10)$$

\noindent ii) We divide the interval $[0,x]$ into $N$ parts and transport
$g(\xi,y)$ from one point to the next using the first order approximation

$$g(\xi+\epsilon,y)\simeq g(\xi,y)+\epsilon O_y g(\xi,y) \qquad
(\epsilon=x/N)$$

By repeatedly doing this, we get:

$$\eqalign{g(\epsilon,y)&\simeq f(y)+\epsilon O_y f(y)\cr
g(2\epsilon,y)&\simeq g(\epsilon,y)+\epsilon O_y g(\epsilon,y)=
f(y)+2\epsilon O_yf(y)+\epsilon^2O^2_yf(y)\cr
\vdots&\cr
g(N\epsilon,y)&\simeq g((N-1)\epsilon,y)+\epsilon O_y g((N-1)\epsilon,y)=
\sum_{n=0}^N {N\choose n}\epsilon^n O^n_yf(y)\cr}$$

Thus, using this approach we get the expansion

$$g(x,y)=\lim_{N\to\infty}\sum_{n=0}^N{N!\over(N-n)!N^n}{x^n\over n!}
O^n_yf(y)\eqno(11)$$

To apply this to the expansion of a function around a point, we just need
to realize that the function $g(x,y)\equiv f(x+y)$ is determined by the
equations

$${\partial g\over\partial x}(x,y)={\partial g\over\partial y}(x,y)\qquad
;\qquad g(0,y)=f(y)$$

\noindent which is the former situation with $O_y=\partial/\partial y$.
Substituting $x$ by $x-x_0$ and $y$ by $x_0$ in (10) and (11) we get the
expressions

$$\eqalignno{i)\qquad f(x)&=\lim_{N\to\infty}\sum_{n=0}^N{f^{(n)}(x_0)
\over n!}(x-x_0)^n&(12)\cr
ii)\qquad f(x)&=\lim_{N\to\infty}\sum_{n=0}^N{N!\over(N-n)!N^n}
{f^{(n)}(x_0)\over n!}(x-x_0)^n&(13)\cr}$$

Term by term, (12) and (13) coincide as $N\to\infty$, but not
uniformly so, and as a result (13) has certain advantages over (12);
for example, for the function $f(x)=1/(1+x)$, (12) converges in the
interval $[0,1]$ while (13) converges in $[0,a]$ where $a$ is the
solution of $log(a)-1/a=1$, ($a\simeq 3.59$). We therefore may wonder
whether our expansion based on ($7'$), ($8'$) already has improved
convergence properties over conventional perturbation theory.
\smallskip

However, let us go further to improve our expansion by the exploitation
of the scaling arbitrariness. As is well known, the eigenfunction
corresponding to a given energy
level, quickly vanishes outside the classically allowed region. Thus,
the distance between the classical turning points of the potential is a
measure of the length scale characteristic of that eigenstate. This scale
can be preserved for all couplings by doing a scaling
transformation which depends on $g$.
\smallskip

The scaling operator is given in the coordinate representation by

$$U(\lambda)\Psi(x)=\sqrt{\lambda}\Psi(\lambda x)\eqno (14)$$
\noindent and transforms the position and momentum operators in this way:

$$U^{-1}(\lambda)XU(\lambda)=X/\lambda \qquad U^{-1}(\lambda)
PU(\lambda)=\lambda P \eqno(15)$$

The Hamiltonian transformed by the operators $U(\lambda(g))$ is then

$$\tilde H(g)=\lambda^{-2}(g)U^{-1}(\lambda(g))H(g)U(\lambda(g))=
{P^2\over 2m}+\lambda^{-2}(g) V(g,X/\lambda(g))\eqno(16)$$

\noindent and $\lambda(g)$ can then be chosen to maintain the scale
corresponding to the $n$th eigenstate by solving the equations

$$E_n(0)=V(0,x_0) \qquad\qquad E_n(0)=\lambda^{-2}(g)V(g,x_0/\lambda(g))
\eqno(17)$$

The first equation determines the classical turning point $x_0$ of the
unperturbed potential for the energy $E_n(0)$ and the second equation
says that $\lambda(g)$ should be chosen by demanding that the classical
turning point for the perturbed potential at the same energy $E_n(0)$ is
again $x_0$. The coupling $g$ can, of course, be conveniently
reparameterized to make sure that equal increments of the coupling
correspond to similar increases of the interaction in the rescaled
Hamiltonian $\tilde H$.
\smallskip

The perturbative expansion for the $n$th state associated to $\tilde H(g)$
will clearly be better than that associated to $H(g)$: $\tilde\Psi_n
(g_f)$ and $\tilde\Psi_n(0)$ have the same number of nodes and are
confined to the same region. In this way $(\tilde\Psi_n(0),\tilde\Psi_n
(g_f))$ is maximized and the terms $(\tilde\Psi_k(0),\tilde\Psi_n(g_f))$
$(k\not= n)$, which express the strength of the perturbation, are minimized.
The eigenvector and eigenvalue corresponding to $H(g_f)$ can then be
immediately obtained from $\tilde\Psi_n(g_f)$, $\tilde E_n(g_f)$ by

$$\eqalignno{\Psi_n(g_f)&=U(\lambda(g_f))\tilde\Psi_n(g_f)&(18)\cr
E_n(g_f)&=\lambda^2(g_f)\tilde E_n(g_f)&(19)\cr}$$

\noindent as follows from (16).
\medskip

What it has been done so far is just setting the appropriate scale
for the $n$th eigenstate. This would correspond in Field Theory to set
the $\mu$ scale equal to the energy at which we are calculating. At
this stage, both the ordinary Taylor expansion and our alternative
expansion could be applied to the rescaled Hamiltonian $\tilde H(g)$. From
here on, we will suppose that we are studying a particular
eigenstate and that we have already done the mentioned {\it global} rescaling
of the Hamiltonian and dropped the tildes from the rescaled quantities.
\smallskip

As one can see in (7) or ($7'$), all energy scales participate in
the calculation of a certain eigenstate and the effect of the perturbation
on each of these energies, should be calculated at their appropriate scales.
It is in taking care of this fact where our alternative perturbative expansion
is much more convenient. Its advantage is that it merely consists in repeating
many times a very simple routine: first order of perturbation theory.
Thus, the scaling arbitrariness can be expressed in a single step very
simply, and this new algorithm repeated for every step. Instead, in a
Taylor expansion, the $N$th order is something very difficult to interpret.
\smallskip

Obviously, the way to exploit the scaling arbitrariness at a single step
is just to do, for every eigenstate and for the elementary increment of the
coupling, the global rescaling that we carried out for our selected state
and the entire range of the coupling.
\smallskip

To be more precise, suppose that $g_1$ and $g_2$ are the initial and final
points of a certain elementary step and that we know $\Psi_l(g_1)$,
$E_l(g_1)$ $\forall l$. The appropriate rescaling of $H(g_2)$ for the $k$th
state is

$$\tilde H(g_2)=\lambda_k^{-2}U^{-1}(\lambda_k)H(g_2)U(\lambda_k)\eqno(20)$$

\noindent where, in analogy with (17), $\lambda_k$ is determined by

$$E_k(g_1)=V(g_1,x_k)\qquad\qquad
E_k(g_1)=\lambda_k^{-2}V(g_2,x_k/\lambda_k)\eqno(21)$$

We now use a first order approximation between $H(g_1)$ and $\tilde H(g_2)$
followed by a scaling transformation between $\tilde H(g_2)$ and $H(g_2)$.
Using (7) and (8) we have the approximations

$$\eqalignno{\tilde\Psi_k(g_2)&\simeq\Psi_k(g_1)+\delta\tilde\Psi_k(g_1)
&(22)\cr
{\rm where}\qquad \delta\tilde\Psi_k(g_1)&\equiv\sum_{l\not= k}
{(\Psi_l(g_1),(\tilde H(g_2)-H(g_1))\Psi_k(g_1))\over E_k(g_1)-E_l(g_1)}
\Psi_l(g_1)\cr
\tilde E_k(g_2)&\simeq E_k(g_1)+(\Psi_k(g_1),(\tilde H(g_2)-H(g_1))
\Psi_k(g_1))&(23)\cr}$$

\noindent which will then be inserted in

$$\eqalignno{\Psi_k(g_2)&=U(\lambda_k)\tilde\Psi_k(g_2)&(24)\cr
E_k(g_2)&=\lambda^2_k\tilde E_k(g_2)&(25)\cr}$$

In the spirit of the more convenient formulas ($7'$) and ($8'$)
we have

$$\tilde\Psi_k(g_2)\simeq(\Psi_k(g_1)+\delta\tilde\Psi_k(g_1))/
\|\Psi_k(g_1)+\delta\tilde\Psi_k(g_1)\|\eqno(26)$$

\noindent which we then substitute in

$$\eqalignno{\Psi_k(g_2)&=U(\lambda_k)\tilde\Psi_k(g_2)&(24)\cr
E_k(g_2)&=(\Psi_k(g_2),H(g_2)\Psi_k(g_2))&(27)\cr}$$

Since $U(\lambda_k)$ is unitary, $\Psi_k(g_2)$ is exactly normalized.
\medskip

Thus, for each state, the approximations are applied to an interaction
that maintains the scale at that particular energy and the change of scale
is carried out by an exact scaling transformation. We have, therefore,
accomplished our goal: in this perturbative expansion, the effect of the
interaction at every energy is calculated at its appropriate scale.
\vfill\eject

\noindent{\bf 3. The anharmonic oscillator: cut-off resummation and scaling.}
\bigskip

In this section, we will investigate the improvements in the perturbation
theory of the anharmonic oscillator when the scaling arbitrariness is
exploited as described in the former section. An additional difficulty
is, however, present in this case: the instability of the system for
negative couplings gives rise to a singularity at $g=0$. We will show that
this singularity is not overcome by the exploitation of scaling and that
can be interpreted as due to the fact that for any order, the perturbative
approximations are inadequate at very high energies. To overcome this
problem, we will introduce a cut-off by limiting the number of states
considered and will provide a criterion for finding the appropriate
order of perturbation that should be associated to every value of the
cut-off to conveniently resum the series. We will finally show how
this order is substantially lower in the case in which the scaling
arbitrariness is fully exploited.
\medskip

We will choose the ground state to carry out all the numerical
analysis. An analogous development could be done for any eigenstate.
We first perform the {\it global} rescaling appropriate for the ground
state explained in section 2.

$$H(g)=P^2+X^2+gX^4 \qquad ([X,P]=i)\eqno (28)$$

$$\eqalignno{\tilde H(g)&=\lambda^{-2}(g)U^{-1}(\lambda(g))H(g)U(\lambda(g))
&(16)\cr
&=P^2+{X^2\over \lambda^4(g)}+g{X^4\over \lambda^6(g)}\equiv P^2+
\alpha X^2+\beta X^4\cr}$$

Using equation (17), we can immediately find that the relationship between
$\alpha$ and $\beta$ is
$$\alpha=1-\beta \eqno (29)$$
Then, the rescaled Hamiltonian (which is naturally parameterized with
$\beta$) is

$$\tilde H(\beta)=P^2+(1-\beta)X^2+\beta X^4\eqno(30)$$

$\beta(g)$ and $\lambda(g)$ are the solutions of

$$\eqalign{\beta&=g/\lambda^6\cr 1-\beta&=1/\lambda^4\cr}\quad\Rightarrow
\quad \eqalign{\beta^3+(g^{-2}-3)\beta^2+3\beta-1&=0\cr
\lambda&=(1-\beta)^{-1/4}\cr}\eqno(31)$$

\noindent and $E_0(g)$ is finally determined by
$$E_0(g)=\lambda^2(g)\tilde E_0(\beta(g))\eqno(32)$$
\noindent where $\tilde E_0(\beta)$ is the
eigenvalue of the rescaled Hamiltonian (30), which will be calculated
perturbatively. It should be noted that while $g$ ranges from zero
to infinity, $\beta(g)$ goes from zero to one. This fact illustrates very
clearly the importance of this rescaling of the Hamiltonian in calculations
for large couplings. From (31), we have that, for large $g$

$$\eqalignno{\beta(g)&=1-{1\over g^{2/3}}+{2\over 3g^{4/3}}+
O({1\over g^2})&(33)\cr
\lambda^2(g)&=g^{1/3}(1+{1\over 3g^{2/3}}+O({1\over g^{4/3}}))&(34)\cr}$$

By inserting these equations in (32) we can determine the asymptotic
behavior of $E_0(g)$:

$$E_0(g)=g^{1/3}\{ \tilde E_0(1)+({\tilde E_0(1)\over3}-{d\tilde E_0
\over d\beta}(1)){1\over g^{2/3}}+O({1\over g^{4/3}})\}\eqno(35)$$

We will later on calculate the coefficient $\tilde E_0(1)$. This is
obviously the most unfavorable quantity to calculate in a perturbative
expansion since it is the most removed from the non interacting theory.
\smallskip

We will study our alternative perturbative expansion for the Hamiltonian
(30) in the two versions
presented in section~2. Expansion I will be the one that does not take
advantage of scaling; eigenvectors and eigenvalues are propagated from
one point to the next using (7)-(8) or ($7'$)-($8'$). Expansion II will
be the one that fully exploits the scaling arbitrariness, propagating
each eigenvector and eigenvalue using the (step and energy dependent)
optimal rescalings described in (22)-(27).
\smallskip

As we pointed out at the beginning of this section, we will need to limit
the number of states of the system in order to resum the series. By limiting
the number of states, we mean that we consider the projection of the
Hamiltonian on the space generated by a finite number of the eigenstates
of the harmonic oscillator. Since the Hamiltonian is invariant under parity,
the parts corresponding to the even and odd states are uncoupled. Thus,
only even states will need to be considered in the calculation of the
ground state energy.
\smallskip

Expansion I can be readily applied to this truncated Hamiltonian. In
fact, it is easy to see that only a finite number of states contribute
to a given order of perturbation and, therefore, beyond this number,
the truncated expansion gives the exact result at that order. For example,
only 8 states contribute at 3rd order and 26 at 4th order.
\smallskip

The limitation of the number of states, brings about a slight
inconvenience in expansion II because the scaling operator does not
map our finite dimensional subspaces into themselves. To overcome this
difficulty it should be noted that everything after equation (20) in
the last section, formally holds if we substitute $U(\lambda_k)$ by
an arbitrary unitary operator. Thus, the obvious thing to do is to
replace $U(\lambda_k)$ by unitary operators that leave invariant our
finite dimensional spaces and coincide with $U(\lambda_k)$ when
the number of states is taken to infinity. Such operators can be obtained
by taking the projection of $U(\lambda_k)$ on the finite dimensional spaces
in the basis formed by our finite number of (even) eigenstates of the
harmonic oscillator and unitarizing it by orthonormalizing Gramm-Schmidt
the columns of this projection starting from the one corresponding to
the lowest eigenstate up.
\smallskip

For a large number of states (large cut-off), these operators will
virtually coincide with $U(\lambda_k)$ at low energy. There will be,
however, some differences for the highest energy states which are necessary
to accommodate the cut-off. Fortunately, as it will be shown, this boundary
effect introduced by the cut-off does not spoil the scaling optimization.
\medskip

The two different ways of carrying out the first order approximations
in both expansions I and II that were presented in section 2, give rise to
two versions of each expansion. We will denote by A those expansions
in which the states are not exactly normalized at each step (eqs. (7)-(8)
and (22)-(25))and by B those in which they are (eqs. ($7'$)-($8'$) and
(26),(24),(27)).
\smallskip

The even eigenstates of the harmonic oscillator will be denoted by
$\mid l\rangle$. In coordinate representation are given by

$$\mid l\rangle =\Bigl({1\over\sqrt{\pi}2^{2l}(2l)!}\Bigr)^{1/2}H_{2l}(x)
e^{-x^2/2}\quad l=0,1,2,\ldots\qquad ((P^2+X^2)\mid l\rangle=
(4l+1)\mid l\rangle)\eqno(36)$$

We will say that the cut-off is equal to $n$ when we limit the number of
states to the set $\mid 0\rangle,\mid 1\rangle,\ldots,\mid n\rangle$.
\smallskip

The numerical calculations that follow were done using `Mathematica' on
a NeXT computer. The programs involved are quite simple: one just needs
to write the algorithm for transporting eigenvectors and eigenvalues from
one intermediate point to the next, and ask the machine to repeat
this algorithm $N$ times.
\smallskip

In tables I and II, we list the values of $E_0(g=1)$ computed respectively
with the A versions of expansions I and II as functions of the order of
perturbation and the cut-off. The exact value  is 1.39235${}^{12}$.
\bigskip\bigskip

\centerline{\bf Table I {\rm (expansion I(A))}}

$$\vbox{\tabskip=0pt\offinterlineskip
\def\tablerule{\noalign{\hrule}}
\halign to 15cm{\strut#&
\vrule#\tabskip=1em plus2em&
\hfil#\hfil&
\vrule#&
\hfil#\hfil&
\vrule#&
\hfil#\hfil&
\vrule#&
\hfil#\hfil&
\vrule#&
\hfil#\hfil&
\vrule#&
\hfil#\hfil&
\vrule#&
\hfil#\hfil&
\vrule#\tabskip=0pt\cr\tablerule
&&$\downarrow$cut-off/order$\rightarrow$&&1&&2&&3&&4&&5&&6&\cr\tablerule
&&2&&1.4671&&1.4177&&1.4060&&1.4025&&1.4009&&1.4000&\cr\tablerule
&&4&&1.4671&&1.4177&&1.4102&&1.4003&&1.3957&&1.3950&\cr\tablerule
&&6&&1.4671&&1.4177&&1.4482&&796.31&&$3.0\times10^9$&&
$9.3\times10^{20}$&\cr\tablerule
&&8&&1.4671&&1.4177&&1.4556&&$3.3\times10^4$&&$7.7\times10^{16}$&&
$5.7\times10^{46}$&\cr\tablerule
&&10&&1.4671&&1.4177&&1.4556&&$4.0\times10^6$&&$3.5\times10^{25}$&&
$1.0\times10^{68}$&\cr\tablerule}}$$
\bigskip\bigskip

\centerline{\bf Table II {\rm (expansion II(A))}}

$$\vbox{\tabskip=0pt\offinterlineskip
\def\tablerule{\noalign{\hrule}}
\halign to 15cm{\strut#&
\vrule#\tabskip=1em plus2em&
\hfil#\hfil&
\vrule#&
\hfil#\hfil&
\vrule#&
\hfil#\hfil&
\vrule#&
\hfil#\hfil&
\vrule#&
\hfil#\hfil&
\vrule#&
\hfil#\hfil&
\vrule#&
\hfil#\hfil&
\vrule#\tabskip=0pt\cr\tablerule
&&$\downarrow$cut-off/order$\rightarrow$&&1&&2&&3&&4&&5&&6&\cr\tablerule
&&2&&1.4671&&1.4178&&1.4061&&1.4025&&1.4009&&1.3999&\cr\tablerule
&&4&&1.4671&&1.4175&&1.4035&&1.3994&&1.3976&&1.3965&\cr\tablerule
&&6&&1.4671&&1.4175&&1.4038&&1.3995&&1.3976&&1.3965&\cr\tablerule
&&8&&1.4671&&1.4175&&1.4039&&1.4087&&1.3982&&1.3965&\cr\tablerule
&&10&&1.4671&&1.4175&&1.4038&&1.8093&&$4.9\times10^8$&&
$3.5\times10^{13}$&\cr\tablerule
&&12&&1.4671&&1.4175&&1.4037&&7.3734&&$1.2\times10^{10}$&&
$-5\times10^{27}$&\cr\tablerule}}$$
\bigskip

It is clear that when we remove the cut-off by taking it to infinity,
we get divergent series in both cases. Thus, the exploitation of
scaling in expansion II is not enough to overcome the divergence
although, as one can see by comparing tables I and II, it tames it
considerably.
\smallskip

To better understand this divergence and the effects of exploiting
scaling, we have listed in tables III and IV (for expansions I and II
respectively) the energies and norms of all states at every intermediate
point for the case ${\rm cut-off}=6$, ${\rm order}=5$, $g=1$
($\beta=0.43015\ldots$).
\vfill\eject

\centerline{\bf Table III {\rm (expansion I(A))}}

$$\vbox{\tabskip=0pt\offinterlineskip
\def\tablerule{\noalign{\hrule}}
\halign to 15cm{\strut#&
\vrule#\tabskip=1em plus2em&
\hfil#\hfil&
\vrule#&
\hfil#\hfil&
\vrule#&
\hfil#\hfil&
\vrule#&
\hfil#\hfil&
\vrule#&
\hfil#\hfil&
\vrule#&
\hfil#\hfil&
\vrule#&
\hfil#\hfil&
\vrule#\tabskip=0pt\cr\tablerule
&&points$\rightarrow$&&0&&1&&2&&3&&4&&5&\cr\tablerule
&&energy(0)&&1&&1.021&&1.034&&1.043&&7.710&&$2.2\times10^9$&\cr\tablerule
&&norm(0)&&1&&1.0005&&1.0008&&1.183&&$1.1\times10^4$
&&$5.4\times10^{12}$&\cr\tablerule
&&energy(1)&&5&&5.623&&5.991&&6.448&&2224&&$7.3\times10^{11}$&\cr\tablerule
&&norm(1)&&1&&1.026&&1.047&&11.71&&$2.0\times10^5$
&&$3.4\times10^{13}$&\cr\tablerule
&&energy(2)&&9&&11.25&&13.27&&27.73&&5346&&$1.9\times10^{11}$&\cr\tablerule
&&norm(2)&&1&&1.187&&1.844&&24.59&&$1.0\times10^5$
&&$1.3\times10^{13}$&\cr\tablerule
&&energy(3)&&13&&17.92&&29.19&&573.7&&$8.8\times10^6$&&
$7.9\times10^{13}$&\cr\tablerule
&&norm(3)&&1&&1.629&&7.237&&716.3&&$2.2\times10^6$
&&$1.7\times10^{14}$&\cr\tablerule
&&energy(4)&&17&&25.62&&75.17&&7013&&$4.6\times10^6$&&
$3.6\times10^{14}$&\cr\tablerule
&&norm(4)&&1&&2.395&&20.52&&557.1&&$4.7\times10^6$
&&$1.0\times10^{14}$&\cr\tablerule
&&energy(5)&&21&&34.35&&166.4&&1791&&$4.1\times10^5$&&
$5.7\times10^{12}$&\cr\tablerule
&&norm(5)&&1&&3.434&&10.65&&152.9&&$5.6\times10^5$
&&$5.5\times10^{13}$&\cr\tablerule
&&energy(6)&&25&&44.12&&234.8&&5697&&$1.0\times10^7$&&
$3.3\times10^{13}$&\cr\tablerule
&&norm(6)&&1&&2.911&&18.36&&812.2&&$1.6\times10^6$
&&$1.2\times10^{14}$&\cr\tablerule}}$$

\centerline{\bf Table IV {\rm (expansion II(A))}}

$$\vbox{\tabskip=0pt\offinterlineskip
\def\tablerule{\noalign{\hrule}}
\halign to 15cm{\strut#&
\vrule#\tabskip=1em plus2em&
\hfil#\hfil&
\vrule#&
\hfil#\hfil&
\vrule#&
\hfil#\hfil&
\vrule#&
\hfil#\hfil&
\vrule#&
\hfil#\hfil&
\vrule#&
\hfil#\hfil&
\vrule#&
\hfil#\hfil&
\vrule#\tabskip=0pt\cr\tablerule
&&points$\rightarrow$&&0&&1&&2&&3&&4&&5&\cr\tablerule
&&energy(0)&&1&&1.021&&1.034&&1.043&&1.050&&1.055&\cr\tablerule
&&norm(0)&&1&&1.0005&&1.0008&&1.001&&1.001&&1.001&\cr\tablerule
&&energy(1)&&5&&5.501&&5.843&&6.115&&6.344&&6.542&\cr\tablerule
&&norm(1)&&1&&1.002&&1.002&&1.002&&1.003&&1.003&\cr\tablerule
&&energy(2)&&9&&10.61&&11.63&&12.42&&13.08&&13.65&\cr\tablerule
&&norm(2)&&1&&1.004&&1.005&&1.005&&1.005&&1.005&\cr\tablerule
&&energy(3)&&13&&16.21&&18.11&&19.67&&21.12&&22.56&\cr\tablerule
&&norm(3)&&1&&1.007&&1.009&&1.012&&1.015&&1.019&\cr\tablerule
&&energy(4)&&17&&22.38&&26.80&&30.27&&34.00&&37.96&\cr\tablerule
&&norm(4)&&1&&1.019&&1.062&&1.076&&1.091&&1.112&\cr\tablerule
&&energy(5)&&21&&32.41&&43.00&&51.39&&60.50&&71.54&\cr\tablerule
&&norm(5)&&1&&1.217&&1.418&&1.458&&1.495&&1.530&\cr\tablerule
&&energy(6)&&25&&51.09&&98.46&&166.1&&233.3&&295.5&\cr\tablerule
&&norm(6)&&1&&1.300&&1.521&&1.541&&1.538&&1.528&\cr\tablerule}}$$
\bigskip

Let us look at the first point. The amount in which the norm of
the states differ from 1 is a measure of the errors made by
perturbation theory in the first step. These errors increase
with energy and are substantially smaller for expansion II showing
very clearly the remarkable improvement introduced by scaling.
\smallskip

As one can see in table III, the divergence in expansion I occurs
because the big errors at high energy are amplified and propagated
to the lower energy states in the subsequent steps, finally reaching
the ground state. The same thing happens in expansion II but for
a larger value of the cut-off ($\simeq10$). The errors of perturbation
theory at a given energy are reduced by increasing the order but then,
we have more steps and the errors from higher energy states better
propagate down to the ground state. Thus, the obvious thing to do is
to keep the cut-off fixed while we increase the order to reduce all
errors as much as we want.
\smallskip

Table V lists the values of $E_0(g=1)$ calculated with expansion I(A)
for a cut-off equal to 6 as we increase the order of perturbation.
\bigskip

\centerline{\bf Table V {\rm (expansion I(A))}}
$$\vbox{\tabskip=0pt\offinterlineskip
\def\tablerule{\noalign{\hrule}}
\halign to 12cm{\strut#&
\vrule#\tabskip=1em plus2em&
\hfil#\hfil&
\hfil#\hfil&
\vrule#&
\hfil#\hfil&
\hfil#\hfil&
\vrule#&
\hfil#\hfil&
\hfil#\hfil&
\vrule#\tabskip=0pt\cr\tablerule
&&order&$E_0(g=1)$&&order&$E_0(g=1)$&&order&$E_0(g=1)$&\cr\tablerule
&&1&1.4671&&7&$1.0\times10^{32}$&&13&1.3934&\cr
&&2&1.4177&&8&$4.7\times10^{38}$&&14&1.3933&\cr
&&3&1.4482&&9&$5.2\times10^{36}$&&15&1.3933&\cr
&&4&796.3&&10&1.3937&&20&1.3930&\cr
&&5&$3.0\times10^9$&&11&1.3936&&25&1.3929&\cr
&&6&$9.3\times10^{20}$&&12&1.3935&&30&1.3928&\cr\tablerule}}$$
\bigskip

At low orders, as we already pointed out, only the lowest energy states
give substantial contributions to $E_0$ and we get decent
approximations. From orders 4 to 9 there are enough steps to let the big
errors at high energy propagate down to the ground state: the numbers
take off. Finally, beyond 10, the order is high enough to curb the errors
at all energies originating excellent approximations.
\smallskip

As we pointed out in the introduction, we are considering a double
expansion: in the cut-off and in the order; and the way in which we
take these quantities to infinity is crucial. If we first remove
the cut-off we get a divergence, while by taking the order to infinity
first and then the cut-off, such divergences are avoided.
\smallskip

Table III shows how the norms of the intermediate states grow very
rapidly causing the energies (which are given by eq. (8)) to take
off as well. Thus, the exact normalization of the states in the B
versions of the expansions should represent a significant improvement.
Table VI lists the values of $E_0(g=1)$ at order 6 as we increase the
cut-off, calculated with expansion I(B) and seems to indicate that there
is no divergence in this case.
\bigskip

\centerline{\bf Table VI {\rm (expansion I(B))}}

$$\vbox{\tabskip=0pt\offinterlineskip
\def\tablerule{\noalign{\hrule}}
\halign to 8cm{\strut#&
\vrule#\tabskip=1em plus2em&
\hfil#\hfil&
\hfil#\hfil&
\vrule#&
\hfil#\hfil&
\hfil#\hfil&
\vrule#\tabskip=0pt\cr\tablerule
&&cut-off&$E_0(g=1)$&&cut-off&$E_0(g=1)$&\cr\tablerule
&&1&1.39819&&7&1.39323&\cr
&&2&1.39654&&8&1.39333&\cr
&&3&1.39351&&9&1.39339&\cr
&&4&1.39299&&10&1.39337&\cr
&&5&1.39312&&11&1.39340&\cr
&&6&1.39338&&12&1.39339&\cr\tablerule}}$$
\bigskip

Whether the B series without cut-off are in fact convergent or not is
somewhat irrelevant because, as one can see in table VI, the best
approximation is obtained for a finite cut-off (4 in this case).
This is due to the fact that for a fixed order, the approximations of
perturbation theory at each step, even though greatly improved and
curbed by normalization, are not valid for high energy states.
\smallskip

As we said, the resummation procedure that we are proposing is to change
the order in which we take the limits in our double expansion: first
take the order to infinity and next the cut-off. However, for a fixed
cut-off, it is only worth increasing the order until the precision
obtained is of the order of the error made by the fact that we are
limiting the number of states. Next, we will give a criterion to
determine when we should stop increasing the order for a fixed cut-off.
\smallskip

In order to do that, we should first know how fast the series for
fixed cut-offs tend to their limits when we increase the order $N$. In the
A expansions, both the states and energies are propagated at each step
with first order of perturbation theory; thus, the error at each step
is of order $(\Delta g)^2\propto 1/N^2$ and the total error will go
like $1/N$. The same applies, in the B expansions, to the eigenstates,
but not to the ground state energy. This is calculated with the formula
$E_0=(\Psi,H\Psi)$. The (approximate) ground state $\Psi$ can be written
as $\Psi=\Psi_0+\chi(1/N)/N$, where $\Psi_0$ is the exact ground state
and $\chi(1/N)$ has a finite limit when $N\to\infty$. Thus,

$$E_0=(\Psi_0,H\Psi_0)+{1\over N}\{(\Psi_0,H\chi(0))+(\chi(0),H\Psi_0)\}+
O(1/N^2)\eqno(37)$$

Since $\Psi_0$ minimizes $E_0$, the term proportional to $1/N$ in (37)
must vanish and we get that the errors in the B expansions go like
$1/N^2$ for large $N$. This fact shows once again the superiority of the
B expansions.
\smallskip

If $E_0(N)-E_0(\infty)$ goes like $1/N$ for the A expansions and like
$1/N^2$ for the B ones, then $Q_0(N)\equiv(E_0(N)-E_0(2N))/(E_0(2N)-
E_0(4N))$ must respectively tend to 2 and 4 for large $N$. In tables VII
and VIII we numerically verify this behavior in the particular case
cut-off=4, $g=1$. Something completely analogous is found for the A and
B versions of expansion II since the use of scaling does not modify
the arguments presented.
\bigskip\bigskip

\centerline{\bf Table VII {\rm (expansion I(A))}}
$$\vbox{\tabskip=0pt\offinterlineskip
\def\tablerule{\noalign{\hrule}}
\halign to 7cm{\strut#&
\vrule#\tabskip=1em plus2em&
\hfil#\hfil&
\hfil#\hfil&
\hfil#\hfil&
\vrule#\tabskip=0pt\cr\tablerule
&&order(N)&$E_0(N)$&$Q_0(N)$&\cr\tablerule
&&50&1.392672024&1.98&\cr
&&100&1.392527424&1.99&\cr
&&200&1.392454389&&\cr
&&400&1.392417667&&\cr\tablerule}}$$
\bigskip\bigskip

\centerline{\bf Table VIII {\rm (expansion I(B))}}
$$\vbox{\tabskip=0pt\offinterlineskip
\def\tablerule{\noalign{\hrule}}
\halign to 7cm{\strut#&
\vrule#\tabskip=1em plus2em&
\hfil#\hfil&
\hfil#\hfil&
\hfil#\hfil&
\vrule#\tabskip=0pt\cr\tablerule
&&order(N)&$E_0(N)$&$Q_0(N)$&\cr\tablerule
&&50&1.392387789&4.10&\cr
&&100&1.392382513&4.04&\cr
&&200&1.392381227&&\cr
&&400&1.392380909&&\cr\tablerule}}$$
\bigskip

We can now give a criterion for when we should stop increasing the order
for a given cut-off. As we said, this should be that the precision
obtained with that order is about the error made by the fact that we are
limiting the number of states.
\vfill\eject

Let $E_0(n,N,g)$ be the $N$th order approximation to $E_0(g)$ when the
cut-off is set equal to $n$. From the knowledge of the asymptotic behavior
we can estimate the precision obtained. For the B versions of the expansions
it will be

$$\epsilon=E_0(n,N,g)-E_0(n,\infty,g)\simeq{4\over3}(E_0(n,N,g)-
E_0(n,2N,g))\eqno(38)$$

On the other hand, a good estimate of the error made by the fact that
we have a finite cut-off is the difference $\Delta=E_0(n-1,N,g)-E_0(n,N,g)
\quad(39)$. Thus, a sensible criterion would be, for example $\epsilon<
\Delta /2$, i.e.

$$E_0(n-1,N,g)-E_0(n,N,g)>{8\over3}(E_0(n,N,g)-E_0(n,2N,g))\eqno(40)$$

\noindent and the appropriate approximation for a cut-off equal to $n$
would finally be:

$$E_0(n,\infty,g)=E_0(n,N,g)-\epsilon\simeq{1\over3}(4E_0(n,2N,g)-E_0
(n,N,g))\eqno(41)$$

\noindent where $N$ is an order that satisfies (40).
\smallskip

As an example, we will apply this technique to compute the coefficient
$\tilde E_0(1)$ of equation (35) that determines the asymptotic behavior of
the ground state energy for large $g$ and which, as we
mentioned, is the most unfavorable
quantity of the anharmonic oscillator to calculate perturbatively. In table
IX we list the smallest order N for which (40) is satisfied and the
corresponding approximation $\tilde E_0(n,1)\equiv(4\tilde E_0(n,2N,1)-
\tilde E_0(n,N,1))/3$
for a cut-off ranging from 3 to 6 and for both expansions I and II (B
versions).
\bigskip\bigskip

\centerline{\bf Table IX}
$$\vbox{\tabskip=0pt\offinterlineskip
\def\tablerule{\noalign{\hrule}}
\halign to 12cm{\strut#&
\vrule#\tabskip=1em plus2em&
\hfil#\hfil&
\vrule#&
\hfil#\hfil&
\hfil#\hfil&
\vrule#&
\hfil#\hfil&
\hfil#\hfil&
\vrule#\tabskip=0pt\cr\tablerule
&&&&expansion\span I(B)&&expansion\span II(B)&\cr\tablerule
&&$n$&&$N$&$\tilde E_0(n,1)$&&$N$&$\tilde E_0(n,1)$&\cr\tablerule
&&3&&6&1.0650&&6&1.0648&\cr
&&4&&7&1.0614&&8&1.0612&\cr
&&5&&19&1.06046&&13&1.06044&\cr
&&6&&100&1.060389&&42&1.060389&\cr\tablerule}}$$
\bigskip

The coefficient $\tilde E_0(1)$ has been computed by variational methods${}^8$,
the result being $1.060362\ldots$ It is remarkable the high degree
of accuracy that is obtained for a relatively low cut-off. For instance,
for $n=6$ the error made is about $3\times10^{-5}$. This is because we
have taken advantage of scaling: the free and interacting ground
states have the same scale and are, therefore, very similar; the small
difference between them can then be very well approximated by a linear
combination of a few higher energy harmonic states.
\smallskip

Finally, it should be noted how, as expected, expansion II (which
makes full use of scaling) is much more efficient than expansion I.
This is more notable the higher the cut-off is, since it is at high
energies where the optimizing rescalings are larger. As an example,
we compare in Table X the values of $\tilde E_0(8,N,1)$ for expansions I
and II. For instance, we only need to calculate to order 40 in
expansion II to get the same precision as with order 100 in expansion I.
\vfill\eject

\centerline{\bf Table X}
$$\vbox{\tabskip=0pt\offinterlineskip
\def\tablerule{\noalign{\hrule}}
\halign to 8cm{\strut#&
\vrule#\tabskip=1em plus2em&
\hfil#\hfil&
\vrule#&
\hfil#\hfil&
\vrule#&
\hfil#\hfil&
\vrule#\tabskip=0pt\cr\tablerule
&&&&expansion I(B)&&expansion II(B)&\cr\tablerule
&&$N$&&$\tilde E_0(8,N,1)$&&$\tilde E_0(8,N,1)$&\cr\tablerule
&&20&&1.0611724&&1.0605232&\cr
&&40&&1.0605931&&1.0604097&\cr
&&60&&1.0604749&&1.0603898&\cr
&&80&&1.0604311&&1.0603830&\cr
&&100&&1.0604105&&1.0603799&\cr\tablerule}}$$
\bigskip

\noindent{\bf Summary}
\bigskip

We have introduced an alternative perturbative expansion which is
very convenient for investigating the contributions of the various
energy scales in the calculation of an eigenvalue and which can be
very naturally improved by carrying out energy-dependent rescalings.
We have identified the divergences in the expansions for the anharmonic
oscillator as due to the fact that the errors made by perturbation
theory in this case cannot be uniformly bounded for all energies.
To overcome this problem, we have introduced an energy cut-off
giving rise to a double expansion in both the order and the range of
energies considered. Finally, we have provided a criterion for how
this limit should be taken in order to get the correct physical
results.
\bigskip\medskip

\noindent{\bf Acknowledgements}
\bigskip

I would like to thank Mark Evans for providing stimulating discussions
and very valuable comments to the original manuscript of this paper. I also
thank N.N.~Khuri, G.B.~West and A.S.~Wightman for discussions
and helpful remarks and Mark Doyle for very generously letting me use
his wonderful NeXT computer.
\bigskip

\noindent{\bf References}
\bigskip

\noindent${}^1$M. Gell-Mann and F.E. Low, {\it Phys. Rev.} {\bf95,}
1300 (1954).
\smallskip

\noindent${}^2$P. Ramond, {\it Field Theory: A Modern Primer},
Addison-Wesley, 1989.
\smallskip

\noindent${}^3$D.J. Gross and F. Wilczek, {\it Phys. Rev. Lett.} {\bf 30,}
1343 (1973).
\smallskip

\noindent${}^4$H.D. Politzer, {\it Phys. Rev. Lett.} {\bf 30,} 1346 (1973).
\smallskip

\noindent${}^5$J.M. Prats, {\it Nucl. Phys.} {\bf B387,} 97 (1992).
\smallskip

\noindent${}^6$J.D. Bjorken and S.D. Drell, {\it Relativistic
Quantum Fields}, McGraw-Hill, 1965.
\smallskip

\noindent${}^7$C.M. Bender and T.T. Wu, {\it Phys. Rev.} {\bf 184,}
1231 (1969).
\smallskip

\noindent${}^8$B. Simon, {\it Annals of Phys.} {\bf 58,} 76 (1970).
\smallskip

\noindent${}^{9}$C.M. Bender and T.T. Wu, {\it Phys. Rev.} {\bf D7,}
1620 (1973).
\smallskip

\noindent${}^{10}$S. Graffi, V. Grecchi and B. Simon, {\it Phys. Lett.}
{\bf 32B,} 631 (1970).
\smallskip

\noindent${}^{11}$J. Zinn-Justin, {\it Phys. Rep.} {\bf 70,} 2 (1981).
\smallskip

\noindent${}^{12}$A. Galindo and P. Pascual, {\it Mecanica Cuantica},
Alhambra, 1978.
\bye